\documentclass[12pt]{article}
\pdfoutput=1

\usepackage{breqn}
\usepackage{putex}
\usepackage{graphicx}
\usepackage{caption}
\usepackage{amsmath}
\usepackage{array}
\usepackage{subcaption}
\usepackage{epstopdf}
\usepackage{enumerate}
\usepackage{cite}
\usepackage{youngtab}
\usepackage{tensor}
\usepackage{slashed}
\usepackage[utf8]{inputenc}
\usepackage{rotating}
\usepackage{bigfoot}
\usepackage[
      colorlinks=true,
      linkcolor=blue,
      urlcolor=blue,
      filecolor=black,
      citecolor=red,
      ]{hyperref}

\newcommand{\abs}[1]{\left\lvert #1 \right\rvert}

\newcommand {\be} {\begin {equation}}
\newcommand {\ee} {\end {equation}}

\newcommand {\bes} {\begin {equation*}}
\newcommand {\ees} {\end {equation*}}

\newcommand{\es}[2] {\begin{equation} \label{#1} \begin{split} #2 \end{split} \end{equation}}

\newcommand{\cA}{{\mathcal A}}

\newcommand{\beq}{\begin{equation}}
\newcommand{\eeq}{\end{equation}}

\def\ie{\begin{equation}\begin{aligned}}
\def\fe{\end{aligned}\end{equation}}

\numberwithin{equation}{section}

\def\<{\langle}
\def\>{\rangle}

\newcommand{\iu}{{i}}

\newcommand{\lb}{\left( }
\newcommand{\rb}{ \right) }

\usepackage{bbm}

\begin{document}

\preprint{}

\institution{weizmann}{Department of Particle Physics and Astrophysics, Weizmann Institute of Science, Rehovot, Israel}

\title{Squashing, Mass, and Holography for 3d Sphere Free Energy}

\authors{Shai M. Chester\worksat{\weizmann}, Rohit R. Kalloor\worksat{\weizmann}, and Adar Sharon\worksat{\weizmann}}

\abstract{
We consider the sphere free energy $F(b;m_I)$ in $\mathcal{N}=6$ ABJ(M) theory deformed by both three real masses $m_I$ and the squashing parameter $b$, which has been computed in terms of an $N$ dimensional matrix model integral using supersymmetric localization. We show that setting $m_3=i\frac{b-b^{-1}}{2}$ relates $F(b;m_I)$ to the round sphere free energy, which implies infinite relations between $m_I$ and $b$ derivatives of $F(b;m_I)$ evaluated at $m_I=0$ and $b=1$. For $\mathcal{N}=8$ ABJ(M) theory, these relations fix all fourth order and some fifth order derivatives in terms of derivatives of $m_1,m_2$, which were previously computed to all orders in $1/N$ using the Fermi gas method. This allows us to compute $\partial_b^4 F\vert_{b=1}$ and $\partial_b^5 F\vert_{b=1}$ to all orders in $1/N$, which we precisely match to a recent prediction to sub-leading order in $1/N$ from the holographically dual $AdS_4$ bulk theory.
}
\date{}

\maketitle

\tableofcontents

\section{Introduction}
\label{intro}

The free energy $F(b,m)$ for a quantum field theory placed on the $d$ dimensional squashed sphere $S^d_b$ and deformed by a mass $m$ is one of the few quantities that can be computed exactly in interacting theories. For a rank $N$ supersymmetric gauge theory, supersymmetric localization has been used to compute $F(b,m)$ in terms of an $N$ dimensional matrix model integral for 2d $\mathcal{N}=(2,2)$ \cite{Gomis:2012wy}, 3d $\mathcal{N}=2$ \cite{Kapustin:2009kz,Hama:2011ea}, 4d $\mathcal{N}=2$ \cite{Pestun:2007rz,Hama:2012bg}, and 5d $\mathcal{N}=1$ \cite{Imamura:2012xg,Imamura:2012bm} theories. The massless theory on the round sphere, i.e. $m=0,b=1$, typically flows in the IR to a conformal field theory. At large $N$ the CFT is often\footnote{This is generically the case for supersymmetric gauge theories with matrix degrees of freedom. For theories with vector degrees of freedom, the large $N$ limit is holographically dual to weakly coupled higher spin theory on $AdS_{d+1}$, see \cite{Giombi:2016ejx} for a review.} dual to weakly coupled supergravity on $AdS_{d+1}$ \cite{Maldacena:1997re}, while turning on mass and squashing in the CFT corresponds to suitably deforming the bulk away from $AdS_{d+1}$. One can then study the weakly coupled gravity theory using the large $N$ CFT either by directly comparing the deformed theories at finite $m,b$, or by using the small $m,b$ expansion in the CFT to constrain correlation functions on flat space, which are then holographically dual to scattering in undeformed $AdS_{d+1}$. For either method, it is crucial to know the explicit large $N$ expansion of $F(b,m)$, not just the matrix model integral given by localization.

This work will focus on $F(b;m_I)$ for the 3d $\mathcal{N}\geq6$ ABJ(M) CFTs \cite{Aharony:2008ug,Aharony:2008gk} with gauge group $U(N)_{k}\times U(N+M)_{-k}$ and Chern-Simons level $k$. Like any 3d $\mathcal{N}=6$ SCFT, ABJ(M) has an $SO(6)_R$ symmetry and a $U(1)$ global symmetry \cite{Bashkirov:2011fr}, so that from the $\mathcal{N}=2$ perspective the theory has an $SO(4)\times U(1)$ flavor symmetry. The theory can then be deformed by three real masses $m_I$, where $m_2,m_3$ correspond to Cartans of the $SO(4)$ and $m_1$ to the Cartan of the $U(1)$. The free energy $F(b;m_I)$ on the squashed sphere in the presence of these masses was computed using localization in terms of an $N$ dimensional matrix model integral in \cite{Hama:2011ea}. For the massless round sphere $F(1;0)$, \cite{Marino:2011eh} showed that this matrix model could be understood as a free Fermi gas with a nontrivial potential, which allowed $F(1;0)$ to be explicitly computed to all orders in $1/N$. This Fermi gas method was then extended to $F(1;m_1,m_2,0)$ (or $F(1;m_1,0,m_3)$) \cite{Nosaka:2015iiw}, and to $F(\sqrt{3};0)$ (or $F(1/\sqrt{3};0)$) \cite{Hatsuda:2016uqa}. For more general $b,m_I$, however, the matrix model takes a more complicated form that is not amenable to this technique. In this work, we will use methods inspired from the Fermi gas approach to derive the exact relation
\es{introResults}{
F(b;m_1,m_2,i\frac{b-b^{-1}}{2})=F(1;\frac{b^{-1}(m_1+m_2)+b(m_1-m_2)}{2},\frac{b^{-1}(m_1+m_2)-b(m_1-m_2)}{2},0)\,,
}
where the RHS is now related to the round sphere expression $F(1;m_1,m_2,0)$ that was computed to all orders in $1/N$. We can then expand both sides around the massless round sphere to derive infinite constraints between $m_I,b$ derivatives at each order. For instance, we find that all combinations of 4 derivatives of $m_I,b$ can be written in terms of the quantities
\es{independents}{
\partial_{m_\pm}^4 F\,,\qquad \partial_{m_+}^2\partial_{m_-}^2 F\,,\qquad \partial_{m_\pm}^2 F\,,\qquad {\Green\partial_{m_2}^2\partial_{m_3}^2 F}\,,\qquad {\Red\partial_{m_\pm}^3\partial_{m_\mp} F}\,,\qquad {\Red\partial_{m_+}\partial_{m_-} F}\,, 
}
all evaluated at $m_I=0$ and $b=1$, where $m_\pm\equiv m_2\pm m_1$ (or $m_\pm\equiv m_3\pm m_1$).\footnote{All $m_I,b$ derivatives of $F$ that are considered in this work are assumed to be evaluated at $m_I=0$ and $b=1$.} The quantities in red have an odd number of $m_\pm$ derivatives and are pure imaginary, the quantity $\partial_{m_2}^2\partial_{m_3}^2F$ in green is generically complex, while the remaining quantities in black are always real. For parity preserving ABJ(M) theories, which includes all $\mathcal{N}=8$ theories, $\partial_{m_\pm}^3\partial_{m_\mp} F$ in fact vanishes, while $\partial_{m_+}\partial_{m_-} F$ vanishes for ABJM theory.\footnote{For ABJ theory with unequal rank, generically $\partial_{m_+}\partial_{m_-} F$ is nonzero, and is related to a choice of background Chern-Simons level \cite{Kapustin:2010mh,Closset:2012vg}.} The ABJ(M) theories have $\mathcal{N}=8$ supersymmetry when $k=1,2$, in which case the $SO(6)_R \times U(1)$ global symmetry is promoted to $SO(8)_R$. As a result $\partial_{m_2}^2\partial_{m_3}^2F$ is related to the other real non-vanishing quantities, which are all written as derivatives of $m_\pm$, or equivalently as derivatives of $F(1;m_1,m_2,0)$. Since $F(1;m_1,m_2,0)$ was computed to all orders in $1/N$ in \cite{Nosaka:2015iiw}, we thus have all orders in $1/N$ expressions for all combinations of 4 derivatives of $m_I,b$ in $\mathcal{N}=8$ ABJ(M) theories. We can similarly relate certain higher order derivatives such as $\partial_b^5F$ in terms of $m_\pm$ derivatives, so that they too can be computed to all orders in $1/N$.

We can then compare these all orders in $1/N$ results for $F(b,m_I)$ to the holographic dual of $U(N)_{k}\times U(N+M)_{-k}$ ABJ(M) theory, which for large $N$ and fixed $M,k$ is dual to weakly coupled M-theory on $AdS_4\times S^7/\mathbb{Z}_k$, while for large $N,k$ and fixed $M,\lambda\equiv N/k$ and then large $\lambda$ is dual to weakly coupled Type IIA string theory on $AdS_4\times \mathbb{CP}^3$.\footnote{Note that the finite value of $M$ does not appear to any perturbative order in these holographic descriptions.} The first way we do this is to directly compare $F(b,m_I)$ to the renormalized on-shell action in the $AdS_4$ theory dual to these deformations. The leading large $N$ term corresponds to the action of $\mathcal{N}=8$ gauged supergravity on $AdS_4$ \cite{deWit:1982bul} evaluated on solutions to the equations of motion that preserve the suitable symmetry group of the CFT deformation, which for $m\neq0$ or $b\neq1$ breaks the amount of supersymmetry to $\mathcal{N}=2$ while preserving certain abelian flavor groups. These solutions were matched to $F(b,m_I)$ at leading order in large $N$ for nonzero $m_I$ in \cite{Freedman:2013oja} and nonzero $b$ in \cite{Martelli:2011fu}. The sub-leading $1/N$ corrections to $F(b,m_I)$ correspond to higher derivative corrections to supergravity evaluated on the corresponding solution. The first higher derivative corrections, i.e. the four derivative terms, were recently derived in \cite{Bobev:2020egg} for any minimal $\mathcal{N}=2$ gauged supergravity on $AdS_4$ in terms of two theory dependent coefficients. These coefficients were then fixed for the ABJ(M) M-theory dual at finite $k$ using the large $N$ results for $F(1,0)$ and the coefficient $c_T$ of the stress tensor two-point function, which was computed using $\partial_{m_\pm}^2F$ in \cite{Agmon:2017xes}. The free energy could then be computed on any asymptotically $AdS_4$ solution to sub-leading order in $1/N$, which for the case of squashing gave \cite{Bobev:2020egg}:
\es{bobev}{
F(b;0)=\frac{\pi \sqrt{2k}}{12}\left[\left(b+\frac1b\right)^2\left(N^{\frac32}+(\frac1k-\frac{k}{16})N^{\frac12}\right)-\frac6k N^{\frac12}\right]+O(N^0)\,.
}
We will match this gravity prediction for the sub-leading $N^{\frac12}$ terms to our all orders in $1/N$ expression for $\partial_{b}^4F$ and $\partial_{b}^5F$ in $\mathcal{N}=8$ ABJ(M) theory, i.e. for $k=1,2$. The further sub-leading powers of $1/N$ in our result will allow the coefficients of future higher derivative corrections to supergravity to be similarly fixed. Note that once these higher derivative terms are known, they can be used to compute gravity quantities on any asymptotically $AdS_4$ solution, not just that corresponding to squashing, and can even be used to compute thermodynamic quantities like higher derivative corrections to the the black hole entropy \cite{Bobev:2020egg}, which are much more difficult to compute directly from CFT using holography.

The second way to constrain the dual $AdS_4$ theory using $F(b,m)$ is using the relation between the small $m,b$ expansion of $F(b,m)$ and integrated correlators of the stress tensor multiplet correlator, which is dual to scattering of gravitons on $AdS_4$. In particular, since both $m_I$ and $b$ couple to operators in the stress tensor multiplet for 3d $\mathcal{N}=6$ SCFTs, it should be possible to relate $n$ derivatives of $F(b;m_I)$ evaluated at $m=0,b=1$ to correlators of $n$ stress tensor multiplet operators integrated on $S^3$ \cite{Closset:2012vg}. These integrated constraints were derived for $\partial_{m_\pm}^4 F$, $\partial_{m_+}^2\partial_{m_-}^2 F$, and $\partial_{m_\pm}^2 F$ for $\mathcal{N}=8$ SCFTs in \cite{Agmon:2017xes,Binder:2018yvd} and for $\mathcal{N}=6$ SCFTs in \cite{Binder:2019mpb}. The stress tensor multiplet four point function in ABJ(M) can then be constrained in the large $N$ limit using analyticity, crossing symmetry, and the superconformal ward identities in terms of just a few terms at each order \cite{Zhou:2017zaw,Chester:2018aca}, whose coefficients can then be fixed using the integrated constraints and the large $N$ expressions for derivatives of $F(b;m_I)$. One can then take the flat space limit of this holographic correlator as in \cite{Penedones:2010ue} and compare to the dual quantum gravity S-matrix in flat space, where $1/N$ corrections correspond to higher derivative corrections to supergravity. This program was carried out to sub-leading order in $1/N$ for the M-theory limit in \cite{Chester:2018aca,Binder:2018yvd}, and the Type IIA limit in \cite{Binder:2019mpb}. To go to further orders, one needs to both derive the integrated constraints for the remaining mass and squashing derivatives, as well as the large $N$ expansions of the localization expressions. This paper completes the latter task for all such fourth order derivatives for $\mathcal{N}=8$ ABJ(M), while for $\mathcal{N}=6$ ABJ(M) a large $N$ expansion is still needed for $\partial_{m_2}^2\partial_{m_3}^2F$. 

 The rest of this paper is organized as follows.  In Section~\ref{review}, we review the matrix model expression of $F(b,m)$ for $U(N)_{k}\times U(N+M)_{-k}$ ABJ(M) theory, as well as previous all orders in $1/N$ results from the Fermi Gas method. In Section~\ref{exact}, we derive the exact relation \eqref{introResults} between mass and squashing, and use it to show that all derivatives up to fourth order as well as $\partial_b^5F$ can be written in terms of the invariants shown in \eqref{independents}. In Section~\ref{large} we use these relations as well as the previously derived all orders in $1/N$ expressions for $F(1;m_1,m_2,0)$ to derive all orders in $1/N$ expressions for $\partial_{b}^4F$ and $\partial_{b}^5F$, which we will match to the gravity prediction in \eqref{bobev}. We end with a discussion of our results and future directions in Section~\ref{conc}.  Details of our calculations are given in various Appendices, and an attached \texttt{Mathematica} notebook includes our result for $c_T$ in the large $k$ weak coupling expansion to $O(k^{-14})$.

 \section{The ABJ(M) matrix model}
 \label{review}
 
 We begin by reviewing ABJ(M) theory and localization results for $F(b;m_I)$, including the all orders in $1/N$ results from the Fermi Gas method for the round sphere with only $m_\pm$ turned on. 
 
 In $\mathcal{N}=2$ language, ABJM theory consists of vector multiplets for each $U(N)_k\times U(N+M)_{-k}$ gauge group, as well as four chiral multiplets $Z^A,W_A$ for $A=1,2$ which transform under the gauge groups and the $SU(2)_R\times SU(2)_R\times U(1)$ flavor symmetry as shown in Table \ref{tab1}. Seiberg duality relates different ABJ(M) theories as 
\es{seiberg}{
U(N)_k \times U(N+M)_{-k} \ \ \ &\longleftrightarrow \ \ \ U(N)_{-k} \times U(N+\abs{k}-M)_{k}\,,\\
}
which implies that $M\leq|k|$. Parity then sends $k\to-k$, so the $M=0$ theories can be seen to be parity invariant from the Lagrangian, while Seiberg duality implies that the $k=M,2M$ theories must be parity invariant on the quantum level.

\begin{table}[htpbp]
\begin{center}
\begin{tabular}{|l|c|c|c|}
\hline
 field     & $U(N)\times U(N+M)$    & $SU(2)\times SU(2)$             & $U(1)$  \\
 \hline 
 $Z^A$     & $(\overline{\bf N},{\bf N})$    & $({\bf 2},{\bf 1})$ & $1$   \\
 $W_A$  & $({\bf N},\overline{\bf N})$     &  $({\bf 1},\overline{\bf 2})$   & $-1$    \\
  \hline
\end{tabular}
\end{center}
\caption{Matter content of ABJM theory and their transformations under gauge and flavor symmetries in $\mathcal{N}=2$ language.}
\label{tab1}
\end{table}

The partition function $Z(b;m_I)=e^{-F(b;m_I)}$ on a squashed sphere with squashing parameter $b$ and deformed by masses $m_I$ for the chiral fields can then be computed by assembling the standard $\mathcal{N}=2$ ingredients, as reviewed in \cite{Willett:2016adv}, to get up to an overall $m_I,b$ independent constant:
 \es{PartFunc}{
  Z(b;m_I) &= e^{\frac{i\pi}{12k}(b-b^{-1})^2M(M^2-1)}
   \int   \frac{d^{N+M} \mu \, d^{N} \nu}{N!(N+M)!} e^{i \pi k \left[ \sum_i \nu_i^2 -\sum_a \mu_a^2 \right]} \prod_{a>b} 4 \sinh \left[\pi b(\mu_a - \mu_b)  \right] \sinh \left[\pi b^{-1}(\mu_a - \mu_b)\right] \\
   &{}\times \prod_{i>j} 4 \sinh \left[\pi b(\nu_i - \nu_j)  \right] \sinh \left[\pi b^{-1}(\nu_i - \nu_j)\right] 
    \prod_{i, a} \biggl[s_b\left(\frac{i Q}{4} - (\mu_a - \nu_i + \frac{m_1 + m_2 + m_3}{2}) \right) \\
    &{}\times s_b\left(\frac{i Q}{4} - (\mu_a - \nu_i + \frac{m_1 - m_2 - m_3}{2} ) \right) 
     s_b\left(\frac{i Q}{4} - (-\mu_a + \nu_i +\frac{-m_1 - m_2 + m_3}{2}) \right) \\
     &{}\times s_b\left(\frac{i Q}{4} - (-\mu_a + \nu_i + \frac{-m_1 + m_2 - m_3}{2}) \right) \biggr]\,,
 }
where $Q=b+\frac 1b$, the $\mu_a,\nu_i$ correspond to the Cartans of the two gauge fields with $a=1,\dots,N+M$ and $i=1,\dots,N$, and each chiral field contributes a factor with masses $m_I$ determined by the charge assignments in Table \ref{tab1}, so that $m_1$ corresponds to $U(1)$ and $m_2+m_3,m_2-m_3$ correspond to the Cartans of each factor in $SU(2)\times SU(2)$, respectively. The functions $s_b(x)$ are reviewed in Appendix \ref{special}. The phase factor was computed for generic $\mathcal{N}=2$ supersymmetric gauge theories in \cite{Imbimbo:2014pla} in terms of the topological anomaly, which was given for ABJ(M) theory in \cite{Kapustin:2010mh,Honda:2014npa}. If we restrict to the round sphere with $b=1$, and set $m_3$ (or $m_2$) zero, then the partition function can be simplified using identities in \ref{special} and written in terms of $m_\pm=m_2\pm m_1$ (or $m_\pm=m_3\pm m_1$) as
 \es{ZABJM}{
 & Z({m_+,m_-} )=  \int \frac{d^{N+M} \mu d^{N} \nu}{N!(N+M)!} e^{i \pi k \left[   \sum_i \nu_i^2-\sum_a \mu_a^2 \right]} \\
  &\times \frac{\prod_{a<b} \left( 4 \sinh^2 \left[ \pi (\mu_a - \mu_b) \right] \right) 
    \prod_{i<j} \left( 4 \sinh^2 \left[ \pi (\nu_i - \nu_j) \right] \right) }{\prod_{i, a} \left(
     4 \cosh \left[\pi (\mu_a - \nu_i + m_+/2  ) \right] 
     \cosh \left[\pi (\nu_i - \mu_a + m_-/2  ) \right]  \right) } \,.
 }
As shown in \cite{Kapustin:2010xq,Binder:2020ckj}, the partition function can furthermore be simplified using the Cauchy determinant formula to take the form 
 \es{ZABJM2}{
Z(m_+,m_-)= \frac{e^{-\frac \pi2MN m_-}Z_0}{\cosh^N \frac{\pi m_+}2}&\int d^Ny\prod_{a<b} \frac{\sinh^2\frac{\pi(y_a-y_b)}{k}}{\cosh\left[\frac{\pi (y_a-y_b)}k +\frac{\pi m_+}{2}\right]\cosh\left[\frac{\pi (y_a-y_b)}k -\frac{\pi m_+}{2}\right]}\\
&\times\prod_{a=1}^N\left(\frac{e^{i\pi y_am_-}}{2\cosh\left(\pi y_a\right)}\prod_{l=0}^{M-1}\frac{\sinh\left[\frac{\pi\big(y_a+i(l+1/2)\big)}{k}\right]}{\cosh\left[\frac{\pi\big(y_a+i(l+1/2)\big)}k-\frac{\pi m_+}2\right]}\right)\,,
 }
where for simplicity all of the overall numerical coefficients are included in the factor $Z_0$. Lastly, as shown in \cite{Marino:2011eh,Nosaka:2015iiw,Binder:2020ckj}, one can use the Cauchy determinant formula to further write this partition function as a free Fermi gas with a single body Hamiltonian that depends on $m_\pm$. One can then use standard methods from statistical mechanics to compute $Z(m_+,m_-)$ to all orders in $1/N$ as
  \es{GotZABJM}{
  & Z(m_-,m_+) = e^A C^{-\frac 13} \text{Ai}\left[C^{-\frac 13} (N-B) \right]+(\text{non-perturbative in $N$}) \,,\\
  C &= \frac{2}{\pi^2 k (1 + m_+^2) (1 + m_-^2)} \,, \qquad
   B = \frac{\pi^2 C}{3} - \frac{1}{6k} \left[ \frac{1}{1 + m_+^2} + \frac{1}{1 + m_-^2} \right] - \frac{k}{12}+\frac k2\left(\frac12-\frac Mk\right)^2 \,, \\
  A&= \frac{{\cal A}[k(1 + i m_+)] + {\cal A}[k(1 - i m_+)] +  {\cal A}[k(1 + i m_-)] + {\cal A}[k(1 - i m_-)] }{4}  \,,
 } 
 where the constant map function ${\cal A}$ is given by \cite{Hanada:2012si}
\es{constantMap}{
{\cal A}(k)&=\frac{2\zeta(3)}{\pi^2k}\left(1-\frac{k^3}{16}\right)+\frac{k^2}{\pi^2}\int_0^\infty dx\frac{x}{e^{kx}-1}\log\left(1-e^{-2x}\right)\\
&=-\frac{\zeta(3)}{8\pi^2}k^2+2\zeta'(-1)+\frac{\log\left[\frac{4\pi}{k}\right]}{6}+\sum_{g=0}^\infty\left(\frac{2\pi i}{k}\right)^{2g-2}\frac{4^gB_{2g}B_{2g-2}}{(4g)(2g-2)(2g-2)!}\,,
}
 and in the second line we wrote $\cA$ in the large $k$ expansion \cite{Hanada:2012si}. Note that the all orders in $1/N$ formula only depends on $M$ via the parameter $B$. 

A useful parameterization of ABJ(M) is given by the coefficient $c_T$ of the two-point function of canonically normalized stress-tensors:
 \es{CanStress}{
  \langle T_{\mu\nu}(\vec{x}) T_{\rho \sigma}(0) \rangle = \frac{c_T}{64} \left(P_{\mu\rho} P_{\nu \sigma} + P_{\nu \rho} P_{\mu \sigma} - P_{\mu\nu} P_{\rho\sigma} \right) \frac{1}{16 \pi^2 \vec{x}^2} \,, \qquad P_{\mu\nu} \equiv \eta_{\mu\nu} \nabla^2 - \partial_\mu \partial_\nu \,.
 }
 This quantity is related to the $AdS_4$ Planck length, and so is a more natural expansion parameter in the holographic large $N$ limit than $N$ itself. We can compute $c_T$ in terms of $Z(m_-,m_+)$ as \cite{Chester:2014fya}
  \es{cTABJM}{
c_T=&\frac{64}{\pi^2}\partial^2_{m_\pm}F\,,\\
}
so that it can be written to all orders in $1/N$ using \eqref{GotZABJM}. It turns out that any other quantity computed by taking $m_I,b$ derivatives of $F(b;m_I)$, when expanded at large $c_T$ in either the M-theory or Type IIA limits, becomes independent of $M$. In this sense these limits are blind to parity, which as discussed depends on the value of $M$. One can also check that only even numbers of $m_\pm$ derivatives are nonzero in this limit, and that these quantities are always real.
 
 \section{Exact relation between squashing and mass}
 \label{exact}
 
 We will now derive the relation between mass and squashing shown in \eqref{introResults}. We will then use this result to show that all quartic order $m_I,b$ derivatives of $F(b;m_I)$ can be written in terms of the invariants \eqref{independents}, and that for $\mathcal{N}=8$ ABJ(M) all of these invariants are known to all orders in $1/N$ from previous Fermi gas results. For $\mathcal{N}=6$ ABJ(M), no such all orders result is known yet for $\partial_{m_2}^2\partial_{m_3}^2F$. We will verify all these relations at both small finite $M,N$ and in the large $k$ weak coupling expansion. Note that in general all $m_I,b$ derivatives are evaluated at $m_I=0$ and $b=1$.
 
 We start by setting $m_3=i\frac{b-b^{-1}}{2}$ in \eqref{PartFunc} and using properties of $s_b(x)$ given in Appendix \ref{special} to write the partition function purely in terms of trigonometric functions:
  \es{PartFunc2}{
  Z(b;m_1,&m_2,i\frac{b-b^{-1}}{2}) = e^{\frac{i\pi}{12k}(b-b^{-1})^2M(M^2-1)}
   \int   \frac{d^{N+M} \mu \, d^{N} \nu}{N!(N+M)!} e^{i \pi k \left[  \sum_i \nu_i^2-\sum_a \mu_a^2 \right]} \\
   &\times\prod_{a>b} 4 \sinh \left[\pi b(\mu_a - \mu_b)  \right] \sinh \left[\pi b^{-1}(\mu_a - \mu_b)\right] 
 \prod_{i>j} 4 \sinh \left[\pi b(\nu_i - \nu_j)  \right] \sinh \left[\pi b^{-1}(\nu_i - \nu_j)\right] \\
&\times    \prod_{i, a} \frac{1}{2\cosh(\pi b^{-1} (\mu_a-\nu_i+\frac{m_+}{2}))} \frac{1}{2\cosh(\pi b (\nu_i-\mu_a+\frac{m_-}{2}))}\,,
 }
 where recall that $m_\pm=m_2\pm m_1$. In Appendix \ref{special}, we then perform the standard Fermi gas steps of writing the products of trigonometric functions as Cauchy determinants, introducing auxiliary variables so that the $\mu,\nu$ factorize into gaussian integrals, and finally performing these integrals and rewriting the Cauchy determinant back into the standard form. The result is
 \es{results}{
 Z(b;m_1,m_2,i\frac{b-b^{-1}}{2})=Z(b^{-1}m_+,bm_-)\,,
 }
 where on the RHS we wrote the simplified round sphere partition function defined in \eqref{ZABJM2}, and note that the $b$ dependent phase that appeared in \eqref{PartFunc} is precisely cancelled, so that the RHS of \eqref{results} depends on $b$ only through a rescaling of the masses. We can then simply rewrite $m_\pm$ in terms of $m_1,m_2$ to get \eqref{introResults}. This entire calculation can also be performed with the roles of $m_2$ and $m_3$ switched with the same result on the RHS of \eqref{results}, which is expected since these masses both correspond to the Cartans of the $SO(4)$ part of the flavor symmetry.
 
 We can now expand both sides of \eqref{results} around $m_I=0$ and $b=1$ to derive relations between derivatives of $F(b;m_I)$. The first nonzero relation appears at quadratic order and relates
 \es{quadratic}{
 \partial_b^2F=2\partial_{m_\pm}^2F+2\partial_{m_+}\partial_{m_-}F\,,
 }
 where we used the fact that various single derivatives of $b$ and $m_I$ identically vanish. From the explicit single variable partition function for $Z(m_+,m_-)$ as given in \eqref{ZABJM2}, we see that
 \es{imaginary}{
\bar Z(m_+,m_-)=(-1)^{MN}Z(-m_+,m_-)\,.
 }
 This implies that any odd number of derivatives of $m_+$ is pure imaginary, such as $\partial_{m_+}\partial_{m_-}F$. We thus conclude that 
  \es{quadratic2}{
 \text{Re}\,\partial_b^2F=2\partial_{m_\pm}^2F\,,
 }
 where recall that $\partial_{m_\pm}^2F$ is manifestly real. This relation is expected from the general results of \cite{Closset:2012vg,Closset:2012ru}, which showed that the real part of two derivatives of any parameter that couples to the stress tensor multiplet should be related to $c_T$, where the precise relation in our case was given in \eqref{cTABJM}. 
 
 At cubic order, we similarly find the nonzero relations
 \es{cubic}{
 i\partial_{m_1} \partial_{m_2} \partial_{m_3}F = -4 \partial_{m_\pm}^2F\,,\qquad \partial_b^3 F=-6\partial_{m_\pm}^2F-6\partial_{m_+}\partial_{m_-}F\,.
 }
Since conformal symmetry fixes both three point and two point functions of the stress tensor to be proportional to $c_T$ \cite{Osborn:1993cr}, we therefore expect that all three derivative terms can be written as linear combinations of two derivative terms. The factor of $i$ on the LHS of the first relation follows from the fact that the real cubic casimir invariant for the $SO(4)\times U(1)$ flavor symmetry (from the $\mathcal{N}=2$ perspective) is $im_1m_2m_3$.

At quartic order, the full list of nonzero relations is
\es{resTab}{
\partial_{m_{2,3}}^4F  &=   \qquad\qquad\;\;\,2 \partial_{m_\pm} ^4 F   + 6\partial_{m_+}^2 \partial_{m_-} ^2 F  \qquad\qquad\quad\;\; + 8 \partial_{m_\pm} ^3 \partial_{m_\mp} F \,, \\
\partial_{m_1}^4 F  &=   \qquad\qquad\;\;\,2 \partial_{m_\pm} ^4 F   + 6\partial_{m_+}^2 \partial_{m_-} ^2 F \qquad\qquad\quad\;\; -8 \partial_{m_\pm} ^3 \partial_{m_\mp} F  \,, \\
\partial_{m_1}^2\partial_{m_{2,3}}^2F  &= \qquad\qquad\;\;\,2 \partial_{m_\pm} ^4 F  - 2\partial_{m_+}^2 \partial_{m_-} ^2 F  \,, \\
\partial_{b}^2\partial_{m_{2,3}}^2F  &=  \;\;8 \partial_{m_\pm} ^2 F  \qquad\qquad\qquad\qquad\qquad\, +\partial_{m_2}^2\partial_{m_3}^2F  \,, \\
\partial_{b}^2\partial_{m_{1}}^2F  &= \; \;8 \partial_{m_\pm} ^2 F  +2\partial_{m_\pm}^4F-2\partial_{m_+}^2\partial_{m_-}^2F \,, \\
\partial_{b}^4F  &=  78 \partial_{m_\pm} ^2 F-2 \partial_{m_\pm} ^4 F -6 \partial_{m_+} ^2\partial_{m_-}^2 F+6\partial_{m_2}^2\partial_{m_3}^2F-8 \partial_{m_\pm} ^3\partial_{m_\mp} F-30 \partial_{m_+} \partial_{m_-}  F    \,, \\
}
which are all written in terms of the six invariants in \eqref{independents}. As discussed above, the two invariants $\partial_{m_+}\partial_{m_-}F$ and $\partial^3_{m_\pm}\partial_{m_\mp}F$ are both pure imaginary, since they involve an odd number of derivatives of $m_+$, and they vanish for ABJM theory with equal rank. The one invariant that cannot generically be written as derivatives of $m_\pm$ is $\partial_{m_2}^2\partial_{m_3}^2F$. For $\mathcal{N}=8$ ABJ(M), however, the flavor group $SO(4)\times U(1)$ is enhanced to $SO(6)$, which implies that $\partial_{m_2}^2\partial_{m_3}^2F$ must be related to the other mass derivatives as \cite{Binder:2018yvd}
\es{N8rel}{
\mathcal{N}=8:\qquad\qquad\partial_{m_I}^2\partial_{m_J}^2F=2\partial_{m_+}^4F-2\partial_{m_+}^2\partial_{m_-}^2F\qquad\text{for}\qquad I\neq J\,.
}
In this case, we find that $\partial_{b}^2\partial_{m_{2,3}}^2F =\partial_{b}^2\partial_{m_{1}}^2F $ in \eqref{resTab} as expected. 

At quintic and higher order of $b,m_I$ derivatives, the relation \eqref{results} is not sufficient to write all derivatives in terms of just $m_\pm$ derivatives even for $\mathcal{N}=8$. For instance, at quintic order $\partial_{m_1}\partial_{m_2}^2\partial_{m_3}^2F$ cannot be further simplified. Nevertheless, at quintic order for $\mathcal{N}=8$ we can use \eqref{results} and \eqref{N8rel} to write $\partial_b^5F$ as
\es{5der}{
\partial_b^5F=-660\partial_{m_\pm}^2F-100\partial_{m_\pm}^4F+180\partial^2_{m_+}\partial_{m_-}^2F\,.
}
At higher order $a>5$, we can no longer write $\partial_b^aF$ in terms of just $m_\pm$ derivatives.

We checked all the relations discussed in this section for $U(N+M)_k\times U(N)_{-k}$ at finite $M,N,k$ as well as in the large $k$ weak coupling expansion. For instance, in Table \ref{check} we show $\partial_{m_1}^2\partial_{m_2}^2F$ and $\partial_{m_2}^2\partial_{m_3}^2F$ for the $U(1)_k\times U(2)_{-k}$ theory for $k=1,\dots 10$, as computed from the explicit partition function in \eqref{PartFunc}. As expected, for $k=1,2$ when the theory is $\mathcal{N}=8$, these quantities are identical and real. For all higher $k$, when the theory is only $\mathcal{N}=6$, these quantities are distinct and $\partial_{m_2}^2\partial_{m_3}^2F$ is complex. Finally, in Appendix \ref{largek} we show that $\partial_{m_1}^2\partial_{m_2}^2F$ and $\partial_{m_2}^2\partial_{m_3}^2F$ differ explicitly in the large $k$ weak coupling expansion, which is automatically $\mathcal{N}=6$. We also computed $c_T$ to $O(k^{-14})$ using an efficient algorithm for the weak coupling expansion, which improves the $O(k^{-2})$ result of \cite{Gorini:2020new}.

\begin{table}
\begin{center}
\begin{tabular}{ccc}
$k$ &$ \partial_{m_1} ^2 \partial_{m_2} ^2 F$ & $\partial_{m_2} ^2 \partial_{m_3} ^2 F$ \\
\hline 
 1 & -24.3523 & -24.3523 \\
 2 & -29.2656 & -29.2656 \\
 3 & -33.3494 &$-32.7324\, -1.05679 i $\\
 4 & -36.5672 & $-35.4997\, -2.0899 i$ \\
 5 & -39.0269 &$-37.748\, -2.83569 i $\\
 6 & -40.8938 &$-39.5664\, -3.29976 i $\\
 7 & -42.317 & $-41.0312\, -3.5507 i $\\
 8 & -43.4129 &$-42.2112\, -3.65555 i$ \\
 9 & -44.2669 & $-43.1652\, -3.66535 i $\\
 10 & -44.9409 & $-43.941\, -3.61564 i $\\
 \hline
\end{tabular}
\end{center}
\caption{Explicit free energy derivatives for the $U(1)_k\times U(2)_{-k}$ ABJ theory for $k=1,\dots 10$, where all derivatives as usual are evaluated at $m_I=0$ and $b=1$.}
\label{check}
\end{table}

\section{Large $N$ and holography}
\label{large}

In the previous section, we showed that for $\mathcal{N}=8$ ABJ(M) theory, we can relate $\partial_b^4F$ and $\partial_b^5F$ to derivatives of $F(m_-,m_+)$ using \eqref{resTab} and \eqref{N8rel}. As reviewed in Section \ref{review}, this quantity was computed to all orders in $1/N$ using Fermi gas methods, which implies that we can also compute $\partial_b^4F$ and $\partial_b^5F$ to all orders in $1/N$. For the $U(N)_{k}\times U(N)_{-k}$ theory with finite $k=1,2$, in which case we have $\mathcal{N}=8$ supersymmetry, we find
\es{b4allorders}{
\partial_b^4F=&10 \sqrt{2k} \pi N^{3/2}
-\frac{5 \pi  \left(k^2-16\right)}{4 \sqrt{2}
   \sqrt{k}} \sqrt{N}+39 k^2 \mathcal{A}''(k)-5 k^4 \mathcal{A}^{''''}(k)-9\\
&+\frac{\pi  \left(5 \left(k^2-32\right) k^2+704\right)
  }{384 \sqrt{2} k^{3/2}}\frac{1}{\sqrt{N}}+O(N^{-1})\,,\\
  \partial_b^5F=&-60 \sqrt{2k} \pi N^{3/2}
+\frac{15 \pi  \left(k^2-16\right)}{2 \sqrt{2}
   \sqrt{k}} \sqrt{N}-330 k^2 \mathcal{A}''(k)+50 k^4 \mathcal{A}^{''''}(k)+30\\
&-\frac{5\pi  \left(k^4-32 k^2+448\right)
  }{64 \sqrt{2} k^{3/2}}\frac{1}{\sqrt{N}}+O(N^{-1})\,,
}
while it is straightforward to compute higher orders in $1/N$. The constant map $\mathcal{A}$ was defined in \eqref{constantMap}, and its derivatives can be computed exactly for any integer value of $k$. In particular, for the $k=1,2$ values that are relevant here, we find that
\es{Aprime}{
{\cal A}''(1)&=\frac16+\frac{\pi^2}{32}\,,\qquad\qquad \,\, \,\,{\cal A}''(2)=\frac{1}{24}\,,\\
{\cal A}''''(1)&=1+\frac{4\pi^2}{5}-\frac{\pi^4}{32}\,,\qquad {\cal A}''''(2)=\frac{1}{16}+\frac{\pi^2}{80}\,.
} 
The leading and sub-leading terms in \eqref{b4allorders} exactly match $\partial_b^4F$ and $\partial_b^5F$ as computed from the bulk prediction \eqref{bobev}. The bulk prediction is in fact for any $k$, not just the $k=1,2$ with $\mathcal{N}=8$ supersymmetry that we could compute here. This implies that \eqref{b4allorders} must hold for any value of $k$ up to $O(N^{\frac12})$. As discussed above, in the large $N$ limit it is more natural to expand quantities in terms of $c_T$ than $N$. Using the large $N$ expansion for $c_T$ given by \eqref{cTABJM} and \eqref{GotZABJM}, we find that $\partial_b^4F$ and $\partial_b^5F$ can be expanded to all orders in $1/c_T$ as
\es{b4allorders2}{
\frac{1}{c_T^2}\partial_b^4F&=\frac{15 \pi
   ^2}{32 c_T}+\frac{24 k^2 \mathcal{A}''(k)-5 k^4 \mathcal{A}^{''''}(k)+6}{c_T^2}+4 \left(\frac{6 \pi}{k^2} \right)^{2/3}
   \frac{1}{c_T^{\frac73}}+O(c_T^{\frac83})\,,\\
   \frac{1}{c_T^2}\partial_b^5F&=-\frac{45 \pi
   ^2}{16 c_T}+\frac{-240 k^2 \mathcal{A}''(k)+50 k^4 \mathcal{A}^{''''}(k)-60}{c_T^2}-40 \left(\frac{6 \pi}{k^2} \right)^{2/3}
   \frac{1}{c_T^{\frac73}}+O(c_T^{\frac83})\,.\\
}
Here, the $1/c_T$ corresponds to the tree level supergravity correction, the $1/c_T^2$ corresponds to the 1-loop supergravity correction, and the $1/c_T^{\frac73}$ term corresponds to the tree level $D^6R^4$ correction. Curiously, the $1/c_T^{\frac53}$ correction, which would correspond to the $R^4$ correction, vanishes. 

Finally, we can also consider the limit of large $N,k$ at fixed $\lambda\equiv N/k$ and then large $\lambda$, which is dual to weakly coupled Type IIA string theory on $AdS_4\times \mathbb{CP}^3$. Using the large $k$ expansion of $\mathcal{A}(k)$ given on the second line of \eqref{constantMap}, we find that
\es{b4allorders3}{
\frac{1}{c_T^2}\partial_b^4F=&\frac{1}{c_T}\left[\frac{15 \pi
   ^2}{32 }-\frac{9 \zeta (3)}{32 \sqrt{2} \pi \lambda^{\frac32}}+O(\lambda^{-3})\right]\\
   &+\frac{1}{c_T^2}\left[5+\frac{9
   \zeta (3)}{\pi ^2 \lambda }-\frac{15\zeta (3)}{2 \sqrt{2} \pi ^3\lambda^{\frac32}}+O(\lambda^{-3})\right]+O(c_T^{-3})\,,\\
   \frac{1}{c_T^2}\partial_b^5F=&\frac{1}{c_T}\left[-\frac{45 \pi
   ^2}{16 }+\frac{45 \zeta (3)}{16 \sqrt{2} \pi \lambda^{\frac32}}+O(\lambda^{-3})\right]\\
   &+\frac{1}{c_T^2}\left[-50-\frac{90
   \zeta (3)}{\pi ^2 \lambda }+\frac{75\zeta (3)}{ \sqrt{2} \pi ^3\lambda^{\frac32}}+O(\lambda^{-3})\right]+O(c_T^{-3})\,.\\
}
Here, the $c_T^{-1}$ term corresponds to the tree level supergravity correction, the $c_T^{-1}\lambda^{-\frac32}$ term corresponds to tree level $R^4$, while the various $c_T^{-2}$ terms correspond to 1-loop corrections. Unlike the M-theory expansion in \eqref{b4allorders2}, we find that the $R^4$ correction no longer vanishes. This result can also be compared to future bulk calculations in this background.

\section{Conclusion}
\label{conc}

The main result of this work is the exact relation between the mass and squashing deformed sphere free energy $F(b;m_I)$ given in \eqref{introResults} for all $\mathcal{N}=6$ ABJ(M) theories. This relation implies infinite relations between derivatives of $F(b;m_I)$ evaluated at $m=0,b=1$, such as the fact that all four derivatives can be written in terms of the six quantities listed in \eqref{independents}. For the $\mathcal{N}=8$ ABJ(M) theories, these relations allowed us to compute $\partial_b^4F(b;m_I)$ and $\partial_b^5F(b;m_I)$ to all orders in $1/N$, which at sub-leading order match the prediction given in \eqref{bobev} from M-theory compactified on $AdS_4\times S^7/\mathbb{Z}_k$ and expanded to leading order beyond the supergravity limit \cite{Bobev:2020egg}. Our results provide constraints at further orders in $1/N$ that will allow more higher derivative corrections to supergravity to be derived following the program outlined in \cite{Bobev:2020egg}.

It is instructive to compare the results of this work for $F(b;m_I)$ in ABJ(M) to similar results in 4d $\mathcal{N}=4$ SYM. The free energy $F(b;m;\tau)$ in this theory was computed using localization in \cite{Pestun:2007rz,Hama:2012bg} in terms of an $N$ dimensional matrix model integral that depends on the complexified gauge coupling $\tau$, a single mass $m$, and the squashing $b$, all of which couple to operators in the $\mathcal{N}=4$ stress tensor multiplet. In \cite{Chester:2020vyz}, it was found that all four derivatives of these three parameters can be written in terms of the three invariants
\es{4d}{
\partial_{m}^4F(b;m;\tau)\,,\qquad \partial_{m}^2\partial_b^2F(b;m;\tau)\,,\qquad c\,,
}
where $c$ is the conformal anomaly and the coefficient of the canonically normalized stress tensor two-point function. Recall that for $\mathcal{N}=8$ ABJ(M) theory, we also found that all four derivatives of $F(b;m_I)$ could be written in terms of the three quantities shown in black in \eqref{independents}, where the similarity to 4d becomes even tighter once we use Table \ref{resTab} to exchange $\partial_{m_+}^2\partial_{m_-}^2F$ for $\partial_{m_\pm}^2\partial_{b}^2F$, and we note that the 3d analog of $c$ is $c_T$, which is proportional to the third invariant $\partial_{m_\pm}^2F$. The fact that there are just three independent quartic derivatives for maximally supersymmetric theories in both 3d and 4d is in some sense expected, as in both cases the unprotected $D^8R^4$ term in the large $N$ expansion of the stress tensor correlator can be fixed in terms of four coefficients \cite{Alday:2014tsa,Chester:2018aca}, so if there were four independent quartic derivatives then one could have derived an unprotected quantity from protected localization constraints. Another similarity between 3d and 4d is that in \cite{Minahan:2020wtz} it was shown that for a special value of the mass $F(b;m;\tau)$ obeys
\es{4d2}{
F(b;i\frac{b-b^{-1}}{2};\tau)=F(1;0;\tau)\,,
} 
i.e. it becomes independent of the squashing and mass, just as in \eqref{introResults} we showed that the exact same relation between $m_3$ and $b$ made $F(b;m_I)$ equivalent to the round sphere free energy with $m_3=0$, although in 3d the dependence on $m_3$ and $b$ is now captured by $m_1$ and $m_2$. It would be interesting to find a deeper geometric explanation for this simplification in all theories where $m,b$ both couple to the stress tensor multiplet.

One application of our results that we did not explore in this work is the relation between $n$ derivatives of $m_I,b$ of $F(b;m_I)$ and correlators of $n$ stress tensor multiplets. For $\mathcal{N}=8$ ABJM theory, $\partial_{m_\pm}^4 F$ and $\partial_{m_+}^2\partial_{m_-}^2F$ were related in \cite{Binder:2018yvd} to integrated constraints on the stress tensor four point function, which were used to derive the large $N$ expansion up to order $D^4R^4$ in the bulk language. One further constraint is needed to fix the $D^6R^4$ term,\footnote{This term could equivalently be fixed using the flat space limit relation to the M-theory S-matrix, which has been computed to order $D^6R^4$.} which is the highest order protected term, and it is possible that the integrated constraint from $\partial_b^2\partial_m^2F$ will be sufficient to fix this term. 

For $\mathcal{N}=6$ ABJ(M), from the list of independent quartic derivatives in \eqref{independents}, we expect that there will now be six total independent constraints. Recall that $\partial_{m_+}\partial_{m_-}F$ and $\partial_{m_\pm}^3\partial_{m_\mp}F$ are known to vanish in the Fermi gas expression that describes both the M-theory and Type IIA string theory limits, so we expect just four constraints in these cases. The integrated constraints for $\partial_{m_\pm}^4 F$ and $\partial_{m_+}^2\partial_{m_-}^2F$ were derived in \cite{Binder:2019mpb} and used to fix the stress tensor correlator in both the M-theory and Type IIA limit to order $R^4$. To fix the correlator to order $D^4R^4$ just from CFT results, one would need six constraints, which is probably more than are even in principle independent. On the other hand, if one uses the known Type IIA amplitude in the flat space limit to fix two of these constraints, then just four more constraints are required, which matches the four invariants we found in this work. To complete this program, one would need to derive the large $N$ expansion of $\partial^2_{m_2}\partial_{m_3}^2 F$, which remains unknown for $\mathcal{N}=6$ ABJ(M).\footnote{Recall that for $\mathcal{N}=8$ ABJ(M) this quantity was related by symmetry to derivatives of $m_+$ and $m_-$ which are known to all orders in $1/N$.} It would be nice if the Fermi gas method for $n$-body operators as initiated in \cite{Chester:2020jay} could be used to compute this quantity. At strong coupling, one could also try to compute them at large $N$ and finite $\lambda\equiv N/k$ using topological recursion as was done for Wilson loops and the free energy in \cite{Drukker:2010nc}. Topological recursion for the ABJM matrix model is quite complicated, however, especially for the multi-body operators we consider, so one could instead try to guess the large $N$ and finite $\lambda$ result from the small $\lambda$, i.e. large $k$, weak coupling expansion, which can be computed to very large order using the algorithm introduced in Appendix \ref{largek} of this work. A first step would be guessing the finite $\lambda$ resummation for $c_T$, which we computed to $O(k^{-14})$ in this work. 

One final interesting limit of $\mathcal{N}=6$ ABJ(M) that we have not yet considered is the large $M,k$ limit at fixed $\tilde\lambda=M/k$ and $N$, which is holographically dual to weakly coupled $\mathcal{N}=6$ higher spin theory \cite{Chang:2012kt}. Unlike the M-theory and string theory limits, this limit is sensitive to the value of $M$, and thus to parity. The parity violating quartic invariants shown in red in \eqref{independents} can also be computed in this limit following \cite{Hirano:2015yha}, and could potentially be used to constrain the correlator. This limit was recently considered in the context of the 3d $\mathcal{N}=6$ numerical bootstrap in \cite{Binder:2020ckj}, and will be further discussed in upcoming work.

\section*{Acknowledgments} 

We thank Ofer Aharony, Itamar Yaakov, Zohar Komargodski, Silviu Pufu, Damon Binder, Marcos Marino, Yifan Wang, Alba Grassi, Masazumi Honda, Erez Urbach and Ohad Mamroud for useful conversation, and Ofer Aharony for reading through the manuscript. We also thank the organizers of “Bootstrap 2019” and Perimeter Institute for Theoretical Physics for its hospitality during the course of this work. SMC is supported by the Zuckerman STEM Leadership Fellowship. This work was supported in part by an Israel Science Foundation center for excellence grant (grant number 1989/14).

\appendix

\section{Details of squashed sphere calculation}
\label{special}

Let us start by reviewing the properties of the double sine function $s_b(x)$ (for reviews see for instance \cite{Bytsko:2006ut,Hatsuda:2016uqa}). This function is defined as
\beq
s_b(x)=\exp\left[-\frac{i\pi}{2}x^2-\frac{i\pi}{24}\left(b^2+b^{-2}\right)+\int_{\mathbb{R}+i0}\frac{dt}{4t}\frac{e^{-2itx}}{\sinh\left(bt\right)\sinh\left(t/b\right)}\right]\,,
\eeq
where the integration contour evades the the pole at $t= 0$ by going into the upper half–plane.
This function obeys several identities:
\begin{enumerate}
	\item $s_b^{-1}(x)=s_b(-x)$.
	\item $s_{b^{-1}}(x)=s_b(x)$.
	\item $s_b\left(\frac{ib}{2}-\sigma\right)s_b\left(\frac{ib}{2}+\sigma\right)=\frac{1}{2\cosh(\pi b\sigma)}$.
\end{enumerate}
The last identity is what we used to get \eqref{ZABJM} and \eqref{PartFunc2}.

Next, we will show how \eqref{PartFunc2} is related to \eqref{ZABJM} as in \eqref{results}, by adapting the usual Fermi gas steps for ABJ as discussed in \cite{Honda:2013pea,Honda:2014npa}. We start from a slightly modified version of \eqref{PartFunc2}:
\begin{gather}
\mathcal{Z} \equiv
Z \lb b; \frac{b\ m_+ - b^{-1} m_-  }{2}, \frac{b\ m_+ + b^{-1} m_-  }{2},  \frac{ \iu \lb b - b^{-1} \rb}{2} \rb
	=
		\mathcal{N}_1
					\int  d ^{M+N} \mu  \ d ^N \nu  \ 		
 	e^{\frac{i\pi}{12k}(b-b^{-1})^2M(M^2-1)}  
	 \nonumber \\
		e^{ - \frac{\iu k }{ 4 \pi} \left( \sum_j \mu_j ^2 - \sum_l \nu_l ^2 \right) }
	 	 \times
	 	 \lb \prod _{ j<l} 2 \ \text{sinh} \left(  b \left( \frac{ \mu _j - \mu _l }{2} \right) \right)  
	 		2 \ \text{sinh} \left( b^{-1} \left( \frac{ \mu _j - \mu _l }{2} \right) \right) \rb
			\lb \prod _{ h<g} 2 \ \text{sinh} \left( b \left( \frac{ \nu _h - \nu _g }{2} \right) \right) \right.
	 \nonumber \\ 
	 \left. 2 \ \text{sinh} \left( b^{-1} \left( \frac{ \nu _h - \nu _g }{2} \right) \right) \rb
	 \times
	 		\prod_{r,s} \left[ \frac{1}{ 2 \ \text{cosh} \left( b \left( \frac{ \mu_r - \nu_s }{2} \right) - \frac{\pi m_-}{2} \right)  } 
			\frac{1}{ 2 \ \text{cosh} \left(  b^{-1} \left( \frac{ \mu_r - \nu_s }{2}  \right) + \frac{\pi m_+ }{2} \right) }  
\right]	\,,
\label{eqn:start}
	\end{gather}
where for convenience we changed variables $ \lb \mu , \nu \rb \rightarrow \lb \mu , \nu \rb / \lb 2 \pi \rb $ relative to \eqref{PartFunc2}, and defined the numerical constant
\begin{align}
\mathcal{N}_1
	& = 
		\frac{1}{ \lb 2 \pi \rb ^{M + 2 N } \lb M +  N \rb ! \ N ! }\,.
\end{align}
Our goal is to show that $\mathcal{Z}$ is independent of $b$. Once this is done, plugging in the right values of $m_{\pm}$ will give \eqref{results}. Our first step will be to use the Cauchy determinant formula \cite{Nosaka:2015iiw,Honda:2013pea} to turn the integrand into a product of two determinants. From here, a clever change of integration variables will let us replace one of the determinants with the product of the diagonal elements of the corresponding matrix. We will then express the integrand as a Fourier transform; as is routinely done in Fermi gas derivations. This allows us do the $ \mu , \nu  $ integrals (these are simple Gaussian integrals at this stage), which gives a simple $b$-dependent phase factor that exactly cancels a similar factor \eqref{eqn:start}, thus making $\mathcal{Z}$ independent of $b$. Since the object of our derivation is to exhibit the $b$-independence of (\ref{eqn:start}), we do not need to keep track of any $b$-independent factors that are produced along the way. Hence, we will employ a series of normalization factors $\mathcal{N}_i$ that soak up all such $b$-independent factors.

We now begin the calculation by using the Cauchy determinant formula as given in \cite{Nosaka:2015iiw,Honda:2013pea} to turn \eqref{eqn:start} into
	\begin{align}
\mathcal{Z}	
    & =
    	\mathcal{N}_1 
    	\int d ^{M+N} \mu  \  d ^{N} \nu \   e^{ -\frac{ \iu k }{ 4 \pi } \lb \sum_j \mu^2 _j - \sum _l \nu^2 _l \rb -\frac{M}{2} \sum_r \lb Q \mu_r + \pi m_+ - \pi m_- \rb + \frac{M}{2} \sum_s Q \nu_s + \frac{i\pi}{12k}(b-b^{-1})^2M(M^2-1)}  
\nonumber \\
	& 	\qquad 
		\left[   \text{det} \lb \Theta_{N,l}  \frac{1}{ 2\ \text{cosh} \frac{ b^{-1} \lb \mu_j - \nu_l \rb + \pi m_+ }{ 2} } + \Theta_{l, N+1} e^{ \lb M + N - l + \frac{1}{2} \rb \lb b^{-1} \mu_j + \pi m_+  \rb } \rb \right] 
	\nonumber \\
	& \qquad \qquad
		\left[  \text{det} \lb \Theta_{N,s}  \frac{1}{ 2\ \text{cosh} \frac{ b \lb \mu_r - \nu_s \rb - \pi m_- }{ 2} } + \Theta_{s, N+1} e^{ \lb M + N - s + \frac{1}{2} \rb \lb b \mu_r - \pi m_- \rb } \rb \right] \,,
\end{align}
where $\Theta_{r,s} = \Theta \lb r - s \rb $ is the step function, and the indices $\lb j, l, r, s\rb$ run from $1$ to $ N + M $.
Now we use the following identity (similar to the one in Appendix A of \cite{Matsumoto:2013nya}):
\begin{align}
&  \int d ^{M+N} \mu  \ d^{N} \nu\ \text{det} \lb f \lb( \mu_j , \nu_l \rb \rb \times \text{det} \lb f \lb( \mu_j , \nu_l \rb \rb
	\nonumber \\		
	&
	\qquad \qquad 	\qquad \qquad
	=
	\lb M+ N \rb ! \int d^{M+N} \mu  \ d^{N} \nu \ \lb \prod_{j=1} ^{M+N} f \lb( \mu_j , \nu_j \rb \rb 
		\times \text{det} \lb f \lb( \mu_r , \nu_s \rb \rb
\end{align}
to get:
\begin{align}
\mathcal{Z}  
       & = 
       \mathcal{N}_2
		\sum_{\sigma \in S_{M+N}} \lb -1 \rb^{  \sigma } \int d^{M+N} \mu  \  d^{N} \nu \    e^{ -\frac{ \iu k }{ 4 \pi } \lb \sum_j \mu^2 _j - \sum _l \nu^2 _l \rb -\frac{M}{2} \sum_r \lb Q \mu_r + \pi m_+ - \pi m_- \rb + \frac{M}{2} \sum_s Q \nu_s + \frac{i\pi}{12k}(b-b^{-1})^2M(M^2-1) }
\nonumber \\
	& \qquad 
	\lb \prod_{j=1}^N  \frac{1}{ 2\ \text{cosh} \frac{ b^{-1} \lb \mu_j - \nu_j \rb + \pi m_+ }{ 2} } \rb 
	\lb \prod_{l=N+1}^{M+N} e^{ \lb M + N - l + \frac{1}{2} \rb \lb b^{-1} \mu_l + \pi m_+  \rb } \rb
	\nonumber \\
		& \qquad \qquad 
		\lb \prod_{r = 1}^N   \frac{1}{ 2\ \text{cosh} \frac{ b \lb \mu_{\sigma(r)} - \nu_r \rb - \pi m_- }{ 2} } \rb
		\lb  \prod_{s = N + 1}^{M + N} e^{ \lb M + N - s + \frac{1}{2} \rb \lb b \mu_{\sigma(s)} - \pi m_- \rb } \rb \,,
 	\end{align}
where we've also expanded the remaining determinant term. We now rewrite the integrals as Fourier transforms:
 	\begin{align}
 	\mathcal{Z}
 			& = 	
			\mathcal{N}_3
		\int d^{M+N} \mu  \  d^{N} \nu \ d ^{N} p \ d ^{M+N} q  \
		  e^{ -\frac{ \iu k }{ 4 \pi } \lb \sum_j \mu^2 _j - \sum _l \nu^2 _l \rb -\frac{M}{2} \sum_r \lb Q \mu_r + \pi m_+ - \pi m_- \rb + \frac{M}{2} \sum_s Q \nu_s + \frac{i\pi}{12k}(b-b^{-1})^2M(M^2-1) }
 \nonumber \\
	& \qquad 
	\lb \prod_{j=1}^{N} \frac{ e^{\iu \frac{ p_j \lb b^{-1} \lb \mu_j - \nu_j \rb + \pi m_+ \rb }{ 2\pi }} }{ 2\ \text{cosh}\ \frac{ p_j }{ 2} } \rb
 	\lb  \prod_{l = N + 1}^{M + N} e^{ \lb M + N - l + \frac{1}{2} \rb \lb b^{-1} \mu_l +\pi m_+ \rb } \rb 
 \nonumber \\
		& \qquad \qquad 
		\lb \prod_{r=N+1}^{M+N} \delta \lb 2 \pi \iu \lb M + N - r + \frac{1}{2} \rb + q_r \rb  e^{  - \iu \frac{ q_r m_- \pi }{ 2\pi }} \rb
		\lb \prod_{s=1}^{N} \frac{ e^{ - \iu \frac{ q_s \lb \pi m_- + b \nu_s \rb  }{ 2\pi }} }{ 2\ \text{cosh}\ \frac{ q_s }{ 2} } \rb
 		\nonumber \\
 			& \qquad \qquad \qquad
 			\lb \sum_{\sigma \in S_{M+N}} \lb -1 \rb^{\sigma} \prod_{t=1}^{M+N}  e^{\iu \frac{ q_t \lb  b  \mu_{\sigma(t)} \rb }{ 2\pi }}  \rb\,.
 \end{align}
The $  \mu , \nu  $ integrals are now easy to do and generate some mass-dependent phases, which are absorbed into the definition of $\mathcal{N}_4$. The resulting expression may be massaged into the following form:
\begin{align}
\mathcal{Z}
& = 
	  \lb \prod_{j=N+1}^{M+N} e^{ \frac{ \iu}{ 4 \pi k}  \lb -2\pi \iu \lb M + N -j + 1/2 \rb b^{-1}  + \iu \pi M Q \rb^2  }  \rb 	
	e^{\frac{i\pi}{12k}(b-b^{-1})^2M(M^2-1)}
	\times \mathcal{N}_4 
	  \sum_{\sigma \in S _{M+N}} \lb -1 \rb^{  \sigma } \int  d ^{M+N} q \ d ^{N} p 
\nonumber \\
	&  \qquad 
		\lb \prod_{l=N+1}^{M+N} e^{ \frac{ \iu}{ 4 \pi k} \lb  -4 \iu \pi q_{\sigma(l)} \lb M + N -l + 1/2 \rb  + q_l^2 b^2  + 2 \iu \pi M Q b q_l \rb }  \rb 	
		\lb \prod_{r=1}^{N}  \frac{ e^{\iu \frac{ p_r   }{ 2\pi } \lb \frac{ q_{\sigma(r)} - q_r }{k} + \pi  m_+ \rb } }{ 2\ \text{cosh}\ \frac{ p_r }{ 2} } \rb
		\nonumber \\ 
		&  \qquad \qquad
			\lb \prod_{s=1}^{N} \frac{ e^{- \iu \frac{   q_s m_-  }{ 2 }} }{ 2\ \text{cosh}\ \frac{ q_s }{ 2} } \rb		
			\lb \prod_{t=N+1}^{M+N} \delta \lb 2 \pi \iu \lb M + N - t + \frac{1}{2} \rb + q_t \rb \rb
		\nonumber\,. \\
	\end{align}
We now observe that the $b$-dependent terms under the integral sign are functions of only of those $q_j$'s with $j \in \{ N+1, ... , M + N \}$. These may be pulled out of the integral with the help of the delta functions -- leaving behind an integral that is completely independent of $b$ (and hence absorbed into the definition of $\mathcal{N}_5$):
	\begin{align}
\mathcal{Z}	
	&  = \mathcal{N}_5
	\lb \prod_{j=N+1}^{M+N} e^{ \frac{ \iu}{ 4 \pi k}  \lb \lb -2\pi \iu \lb M + N -j + 1/2 \rb b^{-1}  + \iu \pi M Q \rb^2 - 4 \pi^2 \lb M + N - j  +1/2 \rb^2 b^2 + 4 \pi^2 M Q b \lb M + N - j + 1/2 \rb  \rb } \rb
	\nonumber \\
	&
		\qquad \qquad \qquad \qquad \times   e^{\frac{i\pi}{12k}(b-b^{-1})^2M(M^2-1)}
	\nonumber \\
		& =
		\mathcal{N}_6\
	e^{ - \frac{ \iu \pi }{ k } \lb b^2 + \frac{1}{b^2} - 2 \rb \frac{ M \lb M^2 - 1 \rb}{ 12} } 
	\times   e^{\frac{i\pi}{12k}(b-b^{-1})^2M(M^2-1)}
	\nonumber \\
	& = \mathcal{N}_6\,,
\end{align}
where note in the second line that the $b$-dependent phase in \eqref{eqn:start} has cancelled. We have thus shown that $\mathcal{Z} = Z \lb b; \frac{b\ m_+ - b^{-1} m_-  }{2}, \frac{b\ m_+ + b^{-1} m_-  }{2},  \frac{ \iu \lb b - b^{-1} \rb}{2} \rb  $ is independent of $b$, so we are now free to plug in $b=1$ to get:
\begin{align}
Z \lb b; \frac{b\ m_+ - b^{-1} m_-  }{2}, \frac{b\ m_+ + b^{-1} m_-  }{2},  \frac{ \iu \lb b - b^{-1} \rb}{2} \rb
	& = 
		Z \lb 1; \frac{ m_+ - m_-  }{2}, \frac{m_+ + m_-  }{2},  0 \rb
			\nonumber \\
	& = 
		Z \lb 1; m_1  , m_2,  0 \rb\,.
\end{align}
This result is valid for any value of $m_+ $ and $m_-$, so we can set $ m_+ \rightarrow b^{-1} m_+ $ and $ m_- \rightarrow bm_-  $ to get \eqref{results}.

\section{The large $k$ expansion}
\label{largek}

In this appendix we compute some observables in ABJ(M) in perturbation theory in the CS level $k$. The computation follows standard procedure, for a recent example see \cite{Gorini:2020new}.

Our starting point is the partition function deformed by real masses on the squashed sphere \eqref{PartFunc}. Taking derivatives with respect to masses and the squashing parameter, and setting them to zero, we can define observables in the matrix model.
In general, this procedure should lead to some expectation value in the matrix model of the form 
\begin{equation}
\langle\mathcal{O}\rangle= \int \frac{d^N \nu d^{N+M} \mu }{N !(N+M)!} 
e^{i \pi k \left(\sum_{i}\nu_{i}^2-\sum_{a}\mu_{a}^{2}\right)} Z_{1-\operatorname{loop}}(\nu_{i}, \mu_{a}) \mathcal{O}(\nu_{i},\mu_a)\;.
\end{equation}
where
\begin{equation}
Z_{1-\operatorname{loop}}\left(\nu_{i}, \mu_{b}\right)=\frac{\prod_{i<j} 4 \sinh ^{2}\left[\pi\left(\nu_{i}-\nu_{j}\right)\right] \prod_{a<b} 4\sinh ^{2}\left[\pi\left(\mu_{a}-\mu_{b}\right)\right]}{\prod_{i, a} 4 \cosh \left(\pi\left(\nu_{i}-\mu_{a}\right)\right) \cosh \left(\pi\left(\nu_{i}-\mu_{a}\right)\right)}\;,
\end{equation}
and $\mathcal{O}(\nu_{i},\mu_a)$ is some operator. 

We are interested in computing such objects in perturbation theory in $1/k$. To simplify notation we first perform the change of variables	
\begin{equation}
x_{i}=\pi \sqrt{k} \nu_{i}, \quad y_{a}=\pi \sqrt{k} \mu_{a}\;,
\end{equation}
which brings our expressions to the form
\begin{equation}\label{eq:temp2}
\langle\mathcal{O}\rangle=\int d X d Y 
e^{\frac{i}{\pi} \left(\sum_{i}x_{i}^2-\sum_{a}y_{a}^{2}\right)} f(x,y) \mathcal{O}(x,y)\;,
\end{equation}
where $d X, d Y$ are the Haar measures\footnote{These include the standard Vandermonde determinant factors.}  for $U(N)$ and
\begin{equation}
f(x, y)=\prod_{i<j} \frac{k\sinh ^{2}\left(k^{-\frac12}\left(x_{i}-x_{j}\right)\right)}{\left(x_{i}-x_{j}\right)^{2}} 
\prod_{a<b}\frac{k\sinh ^{2}\left(k^{-\frac12}\left(y_{a}-y_{b}\right)\right)}{\left(y_{a}-y_{b}\right)^{2}} \frac{1}{\prod_{i, a} \cosh ^{2}\left(k^{-\frac12}\left(x_{i}-y_{a}\right)\right)}\;.
\end{equation}
We will normalize all expectation values as $\langle\mathcal{O}\rangle\to \langle\mathcal{O}\rangle/Z_0$ where
\begin{equation}
Z_0 = \int d X d Y 
e^{\frac{i}{\pi} \left(\sum_{i}x_{i}^2-\sum_{a}y_{a}^{2}\right)}\;.
\end{equation}
Note that the expression \eqref{eq:temp2} has the form of an expectation value of $f(x,y)\mathcal{O}(x,y)$ in a matrix model which is the product of two decoupled free Gaussian matrix models.

In order to perform computations, we must start by explicitly expanding $f$ and $\mathcal{O}$ in $1/k$. For example:
\begin{align}
f(x,y)& = 1+\frac1k\left(\frac{1}{3}\left(\sum_{i<j}x_{ij}^{2}+\sum_{a<b}y_{ij}^{2}\right)-\sum_{i,a}(x_i-y_a)^{2}\right)+O(k^{-2})\;,
\end{align}
And similarly for $\mathcal{O}(x,y)$. Plugging these expansions into \eqref{eq:temp2}, it is clear that we will end up with the expectation value of a sum of products of the form $\sum_i x_i^k$ and $\sum_a y_a^l$ for $k,l\in\mathbb{N}$ in a product of Gaussian matrix models. Since the $X$ and $Y$ Gaussian matrix models are decoupled, these expectation values decouple. So it is enough to be able to compute expectation values of the form 
\begin{equation}
\langle\left(\sum_{i_1} x_{i_1}\right)^{\nu_1} \left(\sum_{i_2} x_{i_2}^2\right)^{\nu_2} \left(\sum_{i_3} x_{i_3}^3\right)^{\nu_3}...\rangle=\langle(\text{tr}X)^{\nu_1}(\text{tr}X^2)^{\nu_2}(\text{tr}X^3)^{\nu_3}... \rangle
\end{equation}
in a free Gaussian matrix model, for some integers $\nu_i$, $i=1,2,...$ (and similarly for $y_a$). Here we have used the fact that the $x_i$'s are the eigenvalues of $X$.

We have thus reduced the problem to that of computing expectation values of multi-trace operators in a free Gaussian matrix model.
These expectation values have closed form expressions  \cite{Itzykson:1990zb}, which we now review. Focus on the $X$ matrix model, and consider the computation of the expectation value of the operator 
\begin{equation}
t_{\bar\nu}=(\text{tr}X)^{\nu_1}(\text{tr}X^2)^{\nu_2}(\text{tr}X^3)^{\nu_3}... 
\end{equation}
where $\nu_i\in\mathbb{N}$. 
The expectation value we are computing is explicitly
\begin{equation}
\langle t_{\bar\nu} \rangle=\frac{\int dX t_{\bar{\nu}}(X)e^{-\frac{1}{2\kappa} \text{tr}X^2}}{\int dX e^{-\frac{1}{2\kappa} \text{tr}X^2}}\;.
\end{equation}
for $\kappa=\frac{i \pi}{2}$. 

We begin by defining $2n=\sum_{j\in\mathbb{N}} j\nu_j$ (note that for odd $\sum_{j\in\mathbb{N}} j\nu_j$ this correlator vanishes). We can think of $\bar \nu$ as a partition of $2n$, $\bar \nu=[1^{\nu_1}...(2n)^{\nu_{2n}}]$. Since partitions of $2n$ are in one-to-one correspondence with classes of the permutation group $S_{2n}$, we will use the same notation for both. Thus, for example, the correlator $(\text{tr}X)^{m}$ corresponds to the identity element $1\in S_{m}$, while the correlator $\text{tr}X^{m}$ corresponds to the longest cycle $(12...m)\in S_{m}$.

Denoting by $\chi^Y$ the irreducible characters of $S_{2n}$ with $Y$ denoting a class of $S_{2n}$,\footnote{The character tables can be generated in Mathematica for small $n$. For example, the character table for $S_4$ is obtained using the command FiniteGroupData[``S4", ``CharacterTable"].} the authors of \cite{Itzykson:1990zb} found
\begin{equation}\label{eq:RMT_correlators}
\langle t_{\bar{\nu}}\rangle= (2n-1)!!\kappa^n\sum_Y\frac{\chi^Y(\bar \nu)\chi^Y([2^n])}{\chi^Y([1^{2n}])}\text{ch}_Y(1)\;.
\end{equation}
The sum here is over all classes $Y$ of $S_{2n}$, and $\text{ch}_Y(1)$ is the dimension of the $su(n)$ representation associated with $Y$.\footnote{This can be computed by mapping the classes $Y$ to Young tableaux. The dimension is then given by the ratio of two numbers. To find the numerator, we start by inserting $n$ into the box in the top-left corner, and then filling in the rest of the boxes such that a step to the right increases the number by $1$, while a step down decreases it by $1$. The numerator is then the product of these integers. The denominator is just the usual hook length of the diagram.}
Equation \eqref{eq:RMT_correlators} represents a fast and efficient way of computing many-body correlators in the free Gaussian matrix model, and thus allows us to compute the observables discussed above to high orders.

As an example, we compute $c_T$ to order $k^{-14}$. We can find the operator $\mathcal{O}$ by using equation \eqref{cTABJM}:
\begin{equation}\label{eq:temp1}
c_T=\frac{64}{\pi^2}\partial^2_{m_+}F=-\frac{64}{\pi^2}\left(\frac{Z^{\prime \prime}}{Z}-\left(\frac{Z^{\prime}}{Z}\right)^{2}\right)\;,
\end{equation} 
where primes denote derivatives by $m_+=m_2+m_1$. The derivatives are given by 
\begin{align}
Z^{\prime}=&- \int \frac{d^N \nu d^{N+M} \mu }{N !(N+M)!}
e^{i \pi k \left(\sum_{i}\nu_{i}^2-\sum_{a}\mu_{a}^{2}\right)} Z_{1-\operatorname{loop}}\left(\nu_{i}, \mu_{a}\right) \frac{\pi}{2}\sum_{a,i} \tanh \pi\left(\nu_{i}-\mu_{a}\right)\;, \\
Z^{\prime \prime}=&  \int \frac{d^N \nu d^{N+M} \mu }{N !(N+M)!}
e^{i \pi k \left(\sum_{i}\nu_{i}^2-\sum_{a}\mu_{a}^{2}\right)} Z_{1-\operatorname{loop}}\left(\nu_{i}, \mu_{a}\right)\nonumber \\
& \times \frac{\pi^{2}}{4}\left(\left(\sum_{a, i} \tanh \left(\pi\left(\nu_{i}-\mu_{a}\right)\right)\right)^{2}-\sum_{a,i} \frac{1}{\cosh ^{2}\left(\pi\left(\nu_{i}-\mu_{a}\right)\right)}\right)\label{eq:Zpp}\;.
\end{align}
Note that $Z'$ vanishes since it is odd under $\nu,\mu\to-\nu,\mu$, and so it is enough to compute $Z''$. The corresponding operator $\mathcal{O}$ can be read off from \eqref{eq:Zpp}:
\begin{equation}
\mathcal{O}(\nu,\mu)=\frac{\pi^{2}}{4}\left(\left(\sum_{a, i} \tanh \left(\pi\left(\nu_{i}-\mu_{a}\right)\right)\right)^{2}-\sum_{a,i} \frac{1}{\cosh ^{2}\left(\pi\left(\nu_{i}-\mu_{a}\right)\right)}\right)\,.
\end{equation}
Following the algorithm above, we computed $c_T$ to $O(k^{-14})$. Due to the length of the expression, we give it in an attached \texttt{Mathematica} file.

We will next be interested in computing $\partial_{m_2}^2\partial_{m_3}^2F$ from equation \eqref{independents} in perturbation theory. Specifically, we would like to compare it to another independent quantity, $\partial_{m_1}^2\partial_{m_2}^2F$, and to show that they are not the same in perturbation theory in large $k$ (where we have $\mathcal{N}=6$ SUSY). Explicitly, $\partial_{m_2}^2\partial_{m_3}^2F$ is given by
\begin{equation}\label{eq:XoverZ}
	\partial_{m_2}^2\partial_{m_3}^2F=-\frac{\partial_{m_2}^2\partial_{m_3}^2Z}{Z}+\frac{\partial_{m_2}^2Z}{Z}\frac{\partial_{m_3}^2Z}{Z}\;.
\end{equation}
Using the relation \eqref{introResults}, we find that we can write the first term as
\begin{equation}
\frac{\partial_{m_2}^2\partial_{3}^2Z}{Z}=4\frac{\partial_{m_-}^{2}\partial_{b}^{2}Z}{Z}-\frac{\left(2\partial_{m_+}^{4}-2\partial_{m_+}^{2}\partial_{m_-}^{2}+16\partial_{m_+}^{2}\right)Z}{Z}\;,
\end{equation}
while we find
\begin{equation}
	\frac{\partial_{m_2}^2Z}{Z}=\frac{\partial_{m_3}^2Z}{Z}=2\frac{\partial_{m_+}^2}{Z}+2\frac{\partial_{m_+}\partial_{m_-}Z}{Z}\;.
\end{equation}
We thus have to compute the derivatives $\partial_{m_-}^{2}\partial_{b}^{2}Z$, $\partial_{m_+}^{4}Z$, $\partial_{m_+}^{2}\partial_{m_-}^{2}Z$, $\partial_{m_+}^{2}Z$, $\partial_{m_+}\partial_{m_-}Z$.
First, we write the operators $\mathcal{O}(x_i,y_a)$ corresponding to each term:
\small
\begin{align}
\partial_{m_-}^{2}\partial_{b}^{2}Z\to&\sum_{a,i}\frac{1}{8}\pi^{2}\text{sech}^{2}(g_{s}r_{ai})\left(20g_{s}r_{ai}\tanh(g_{s}r_{ai})-2\left(4g_{s}^{2}r_{ai}^{2}+\pi^{2}+4\right)+3\left(4g_{s}^{2}r_{ai}^{2}+\pi^{2}\right)\text{sech}^{2}(g_{s}r_{ai})\right)\nonumber\\
&+\frac{\pi^{2}}{4}\sum_{b,j}\tanh\left(g_{s}r_{bj}\right)\sum_{a,i}\text{sech}^{2}\left(g_{s}r_{ai}\right)\left(6g_{s}r_{ai}-\left(4g_{s}^{2}r_{ai}^{2}-\cosh\left(2g_{s}r_{ai}\right)+\pi^{2}-1\right)\tanh\left(g_{s}r_{ai}\right)\right)\nonumber\\
& +O_{b}O_{+}(0) \\
\partial_{m_+}^{4}Z\to&  O_+^2(0)-2\partial_mO_+(m)\frac{\pi}{2}\sum_{a,i}\tanh\left(\pi r_{ai}\right)+\partial_m^2O_+(m)\\
\partial_{m_+}^{2}\partial_{m_-}^{2}Z \to &O_+^2(0)\\
\partial_{m_+}^{2}Z\to & O_+(0)\\
\partial_{m_+}\partial_{m_-}Z\to&-\frac{\pi^2}{4} \left(\sum_{a,i}\tanh\left(g_{s}r_{ai}\right)\right)^{2}
\end{align}
\normalsize
where $r_{ai} \equiv x_a-y_i$. Here
\begin{align}
O_+(m)=&\frac{\pi^{2}}{4}\left(\left(\sum_{a,i}\tanh\left(g_{s}r_{ai}+\frac{\pi m}{2}\right)\right)^{2}-\sum_{a,i}\frac{1}{\cosh^{2}\left(g_{s}r_{ai}+\frac{\pi m}{2}\right)}\right)\;,\\
O_b=&\sum_{a<b}\left(2g_{s}x_{ab}\left(\coth\left(g_{s}x_{ab}\right)-g_{s}x_{ab}\text{csch}^{2}\left(g_{s}x_{ab}\right)\right)\right)+\left(x_{ab}\leftrightarrow y_{ij}\right)\nonumber\\
&-\frac{1}{2}\sum_{a,i}\left(4g_{s}^{2}r_{ai}^{2}+2g_{s}r_{ai}\sinh(2g_{s}r_{ai})+\pi^{2}\right)\text{sech}^{2}(g_{s}r_{ai})\;.
\end{align}
Now we can compute the expectation values. 
Plugging these into equation \eqref{eq:XoverZ}, we find 
\begin{dmath}
	\partial_{m_2}^2\partial_{3}^2F=
	-\frac{1}{4} N (M+N) \pi ^4-\frac{1}{6} i N
	(M+N) \pi ^3 \left(-28 M+4 M^3+3 M \pi
	^2\right) \frac{1}{k}+\frac{1}{24} N (M+N) \pi
	^3 \left(-96 M^2 \pi -5 \pi ^3+13 M^2 \pi
	^3+5 M N \pi ^3+5 N^2 \pi ^3\right)
	\frac{1}{k^2}+\frac{1}{36} i N (M+N) \pi ^3
	\left(-232 M \pi ^2-132 M^3 \pi ^2+4 M^5
	\pi ^2-56 M^2 N \pi ^2+8 M^4 N \pi ^2-56
	M N^2 \pi ^2+8 M^3 N^2 \pi ^2+21 M \pi
	^4+15 M^3 \pi ^4+15 M^2 N \pi ^4+15 M N^2
	\pi ^4\right) \frac{1}{k^3}
	+\frac{N (M+N) \pi ^3 }{1440} \left(15360
	\pi ^3+26880 M^2 \pi ^3+2880 M^4 \pi
	^3-19200 M N \pi ^3+1920 M^3 N \pi
	^3-19200 N^2 \pi ^3+5760 M^2 N^2 \pi
	^3+7680 M N^3 \pi ^3+3840 N^4 \pi ^3-1644
	\pi ^5-2545 M^2 \pi ^5-363 M^4 \pi
	^5+2680 M N \pi ^5-717 M^3 N \pi ^5+2680
	N^2 \pi ^5-1753 M^2 N^2 \pi ^5-2072 M N^3
	\pi ^5-1036 N^4 \pi ^5\right) \frac{1}{k^4}\;.
\end{dmath}
We would like to compare this result to $\partial_{m_1}^2\partial_{m_2}^2F$. Using the definitions of $m_\pm$ we find
\begin{equation}
\partial_{m_1}^2\partial_{m_2}^2Z=2\partial_{m_+}^4Z-2\partial_{m_+}^2\partial_{m_-}^2Z\;.
\end{equation}
Using the results above we can compute this as well:
\begin{dmath}
	\partial_{m_1}^2\partial_{m_2}^2F=
	-\frac{1}{4} N (M+N) \pi ^4+\frac{1}{24} N
	(M+N) \pi ^4 \left(-5 \pi ^2+M^2 \pi ^2+5
	M N \pi ^2+5 N^2 \pi ^2\right)
	\frac{1}{k^2}+\frac{N (M+N) \pi ^4  }{2073600}
	\left(-293760
	\pi ^4+396000 M^2 \pi ^4-4320 M^4 \pi
	^4+1267200 M N \pi ^4-168480 M^3 N \pi
	^4+1267200 N^2 \pi ^4-1141920 M^2 N^2 \pi
	^4-1946880 M N^3 \pi ^4-973440 N^4 \pi
	^4\right)\frac{1}{k^4}\;.
\end{dmath}
In general, we find that $\partial_{m_2}^2\partial_{m_3}^2F\neq \partial_{m_1}^2\partial_{m_2}^2F$. For example, for ABJM (where $M=0$), they agree up to order $1/k^3$, with the first difference appearing at order $1/k^4$. Specifically, for ABJM we find
\begin{equation}
	\partial_{m_2}^2\partial_{m_3}^2F-\partial_{m_1}^2\partial_{m_2}^2F=
	-\frac{\pi ^6N^2}{12} 	
	 \left(3 \pi ^2-32\right) \left(N^4-5 N^2+4\right)\frac{1}{k^4} + O\left(\frac{1}{k^6}\right)\;.
\end{equation}

\bibliographystyle{ssg}
\bibliography{squash}

\end{document}